\documentstyle[aps,psfig,preprint] {revtex}
\tightenlines
\begin{document}
\title{Effects of in-medium cross-sections and optical potential on thermal-source formation
in p+$^{197}$Au reactions at 6.2-14.6 GeV/c}

\author{S. Turbide$^{1,2}$,
L. Beaulieu$^1$, P.
Danielewicz$^3$, V.E. Viola$^4$, R. Roy$^1$, K. Kwiatkowski$^{4,}$\footnote[2]{Present address:
Los Alamos National Laboratory, Los Alamos, NM 87545}, W.-C. Hsi,$^{4,}$\footnote[3]{Present address: Rush
Presbyterian St. Lukes Medical Center, Chicago, IL 60612}, G. Wang$^{4,}$\footnote[4]{Present address:
Epsilon Inc., Dallas TX 75240}, T. Lefort$^{4,}$\footnote[5]{Present address:
LPC Caen, 6 Boulevard Mar\'echal Juin, 14050 Caen Cedex, France}, D.S.
Bracken$^{4,}$\footnote[6]{Present address:  Los Alamos National Laboratory, Los Alamos, NM 87545},
H. Breuer$^5$, E. Cornell$^{4,}$\footnote[7]{Present address:  Lawrence
Berkeley National Laboratory, Berkeley, CA 94720}, F. Gimeno-Nogues$^6$,
D.S. Ginger$^{4,}$\footnote[8]{Present address:  Department of
Physics, Cambridge University, United Kingdom}, S. Gushue$^7$, R. Huang$^3$,
R. Korteling$^8$, W.G. Lynch$^3$, K.B. Morley$^9$, E. Ramakrishnan$^6$, L.P.
Remsberg$^7$, D. Rowland$^6$, M.B. Tsang$^3$, H. Xi$^3$ and S.J. Yennello$^6$}

\address{
$^1$D\'epartement de Physique, de G\'enie Physique et d'Optique,
Universit\'e Laval, Qu\'ebec,\\ Canada, G1K 7P4\\
$^2$Department of Physics, McGill University, 3600 University Street, Montreal, Canada H3A 2T8\\
$^3$Department of Physics and NSCL, Michigan State University, East
Lansing, MI 48824 \\
$^{4}$Department of Chemistry and IUCF, Indiana University,
  Bloomington, IN 47405 \\
$^5$Department of Physics, University of Maryland, College Park,
Maryland 20740\\
$^6$Department of Chemistry and Cyclotron Institute, Texas A\&M
University,\\ College Station, TX 77843\\
$^7$Chemistry Department, Brookhaven National Laboratory, Upton, New
York 11973\\
$^8$Department of Chemistry, Simon Fraser University, Burnaby,
BC, Canada V5A IS6\\
$^9$Physics Division, Los Alamos National Laboratory, Los Alamos, NM 87545}

\date{\today}
\maketitle

\begin{abstract}
Effects of in-medium cross-sections and of optical potential on pre-equilibrium
emission and on formation of a thermal source are investigated by comparing the results
of transport simulations with experimental results from the p+$^{197}$Au reaction at 6.2-14.6
GeV/c.  The employed transport model includes light composite-particle production and allows
for inclusion of in-medium particle-particle cross-section reduction and of momentum
dependence in the particle optical-potentials.  Compared to the past, the
model incorporates improved parameterizations of elementary high-energy processes.
The simulations indicate that the majority of energy deposition occurs during the first 25~fm/c
of a reaction.  This is followed by a pre-equilibrium emission and readjustment of system
density and momentum distribution toward an equilibrated system.  Good agreement
with data, on the d/p and t/p yield ratios and on the residue
mass and charge numbers, is obtained at the time of $\sim 65$~fm/c from the start of a reaction, provided
reduced in-medium cross-sections and momentum-dependent optical potentials are employed in the simulations.
By then, the pre-equilibrium nucleon and cluster emission, as well as mean-field readjustments,
drive the system to a state of depleted average density, $\rho/\rho_{0} \sim$ 1/4-1/3 for
central collisions, and low-to-moderate excitation, i.e.\ the region of nuclear liquid-gas phase transition.
\end{abstract}

\vspace {5mm}

\section{INTRODUCTION}
Heating of heavy nuclei with GeV hadrons, and a subsequent
decay of those nuclei, has been the subject of many experiments in the last decade
\cite{beau2,hsi,gold,Avd1,lef1,beau}.  The target-like residues, formed in this way,
primarily decay in the course of statistical and thermal processes including fission,
evaporation\cite{gold} and, for the most highly excited species, multifragmentation\cite{lef2}.
This latter process is of particular interest for its possible relation to the nuclear
liquid-gas phase transition\cite{wang}.  The formation of the residue and reaching an equilibrium, however, are preceded by a fast cascade and pre-equilibrium processes when
energetic nucleons and light clusters are emitted, impacting the equilibration time,
density and excitation energy of the hot residue.

The goal of the present work is to investigate the energy-dissipation process that leads to the formation of highly-excited,
thermal-like residues in reactions of 6.2-14.6 GeV/c protons with $^{197}$Au
nuclei, and to highlight the importance of in-medium cross-sections and
momentum-dependent potentials for the process.  The importance of the momentum
dependence \cite{Jeukenne} for heavy-ion reactions has been demonstrated
in the past by Zhang, Das Gupta and Gale \cite{Zhang}, and by Pan and Danielewicz \cite{Pan},
who have shown that the features of measured collective flow in the reactions could only be reproduced,
in transport simulations, if a momentum-dependent potential was assumed for nucleons, in combination with a low nuclear
incompressibility.  However, it is important to find out whether comparable conclusions can be drawn
on the basis of other observables.  Along similar lines, reduced in-medium cross-sections in transport codes \cite{dan00}
have been needed to explain data on linear momentum transfer.  However, could the reduction be confirmed
by looking at other aspects of the reactions, to generate a consistent interpretation of the
reactions? Pauli-principle effects on intermediate states in scattering, in particular,
can lower the cross-sections \cite{dan00}.

In the incident-momentum range of 6.2-14.6 GeV/c, the energy is deposited in different
processes: hard nucleon-nucleon scattering, excitation of baryonic
resonances and their subsequent decay and reabsorption of some fraction of
the decay pions.  As momentum distributions within the target residue randomize,
pre-equilibrium emissions continue, carrying imprints of the randomization
processes.  mean-fields, together with collisions, determine how the residue
density relaxes.  The coalescence of nucleons into composite particles during
randomization influences the energy dissipation, thereby affecting the excitation
energy ending up in the formed thermal-like residue.  A unique feature of our
analysis is the ability to correlate light-charged particle emission
with the measured excitation energy of the hot nucleus.  In
Ref.~\cite{beau}, it was shown for this data set that the increase
in excitation energy, measured by calorimetry, is correlated with
increased charge and mass loss in the early stage of a reaction.
It needs to be pointed out that the energetic light particles in our data
set correspond to the low-energy component ($<$ 400 MeV) of the "grey particles"
observed in emulsion studies.  Their multiplicity in different reactions has been related to the average
number of hadron-hadron collisions in crossing the nucleus \cite{bus75}.
More recently, Chemakin {\em et al.}\ have shown that the
grey particle multiplicity is correlated with collision centrality\cite{che99}.

A variety of transport models has been employed, in the past, in relating the spectra of
observed particles to the features of an evolving nuclear system in the GeV proton-induced reactions.
The models included the intranuclear cascade \cite{isab,cugn,qgsm,ilji} and transport
models based on the Boltzmann-Uehling-Uhlenbeck (BUU) equation \cite{dan,bauer}.
When the emission of fast particles is of interest, as well as the process of formation of a
thermal residue and its features, a range of problems is encountered in confronting a model with the data.
Among those are the inclusion of all necessary energetic scattering cross-sections, the resonance formation and decay, inclusion of mean-field effects
for the target-like residue, and the complex particle production.
In the present analysis, we rely on the Boltzmann-equation model (BEM) \cite{dan},
that contains appropriate cross-section parameterizations for the high-energy regime,
has a flexibility with regard to the choice of mean-fields importance in residue formation
and describes the production of light composite particles, $^{2}$H, $^{3}$H and $^{3}$He.
The simulations rely on the so-called global ensemble.
The production of the composite particles, in particular, affects the energy dissipation.
An earlier version of this model has been employed by Wang {\em et al.} \cite{wang}
in their analysis of hadron-induced reactions at a few GeV/c.  They found a region of depleted
density in the target nucleus after penetration by the projectile.  The relaxation of the
structure was linked to the multifragmentation of the system.  However, at the time,
the parameterization of high-energy processes largely represented an extrapolation of those appropriate for low
energies and the mean-fields had just a density dependence.  The impact of cluster formation on residue formation was
not studied.  Now, the transport code with more physical features is employed in the same energy
regime to verify if it can reproduce observables related to the process of energy
dissipation. In particular, we are interested in the composite-particle production, prior to
equilibrium, and in the excited-source properties.  We also want to examine the
``apparent'' thermalization time.

In Sec.\ II, assumptions behind the BEM simulations are presented. Section III provides some details
on the experimental set-up and analysis.  Section IV discusses the impact of in-medium cross-sections
and of momentum-dependence of optical potentials on the progress of the p+Au reactions and the formation
of an equilibrated source.  Results from filtered BEM simulations
are compared to the data in Sec.\ V and some additional predictions of the simulations are presented in Sec.\ VI.
In the final section, results from the analysis are reassessed and avenues for further investigations,
in particular of isospin effects, are presented.

\section{Boltzmann-Equation Model}

The transport simulations rely on a set of Boltzmann equations for the
evolution of the hadronic phase-space distribution functions
$f(\overrightarrow{r},\overrightarrow{p},t)$:
\begin{equation}
\label{BUU}
\frac{\partial f}{\partial t} +\nabla_{p}\, \epsilon\cdot\nabla_{r} \, f- \nabla_{r} \, \epsilon\cdot \nabla_{p} \, f
= \left(\frac{\partial f}{\partial t} \right)_{coll} \, .
\end{equation}
Here, $\left(\frac{\partial f}{\partial t}\right)_{coll}$ represents the rate of change of
$f$ caused by collisions and decays and $\epsilon$ is the single-particle energy.
In the case of momentum-independent mean-fields, the single-particle energy is
of the form
\begin{equation}
\label{energy}
\epsilon=\sqrt{p^2+m^2(\rho)}
+ t_z \, a_T \, \frac{\rho_T}{\rho_0} + Z \, \Phi_C \, ,
\end{equation}
where $\rho$ and $\rho_T$ are the scalar baryon and isospin densities \cite{dan00},
$\rho_0 = 0.16 \, \mbox{fm}^{-3}$ is the normal density,
$\Phi_C$ is the Coulomb potential and $t_z$ and $Z$ are, respectively, particle isospin and charge number.
The in-medium density-dependent mass of a particle of mass number $A$ is given by
\begin{equation}
m(\rho)=m+A \, U^\rho \, ,
\end{equation}
with the scalar potential $U^\rho$ parametrized as
\begin{equation}
\label{U=}
U^\rho =\frac{-a\frac{\rho}{\rho_0}+b(\frac{\rho}{\rho_0})^{\nu}}{1+(\frac{\rho}{\rho_0}/2.5)^{\nu-1}} \, .
\end{equation}
The parameter values above are equal, for the soft equation of state, to $a$=187.24 MeV, $b$=102.62 MeV and $\nu$= 1.6339.
The second term in Eq.~(\ref{energy}) is the symmetry term, with $a_T$=97 MeV, while the last term is the Coulomb energy.
In the case of momentum-dependent mean-fields $U^p$, the single-particle energies are
given by
\begin{equation}
\epsilon = \sqrt{p^2+m^2} + A \, U^p
+ t_z \, a_T \, \frac{\rho_T}{\rho_0} + Z \, \Phi_C \, ,
\end{equation}
where, in the local frame,
\begin{equation}
\sqrt{p^2+m^2} + A \, U^p =
m + \int_0^p dp' \, v^* + A
\left[  \rho
\left\langle \int_0^{p_1} dp' \, {\partial v^* \over \partial
\rho} \right\rangle + U^\rho \right] \, .
\end{equation}
Here, $v^*$ is the local velocity of the form
\begin{equation}
\label{v=}
v^* = \frac{p}{ \sqrt{p^2 + m^2
\left/ \left( 1 +  \frac{\rho}{\rho_0} \frac{c}{
(1 + \lambda \, p^2/m^2)^2} \right)^2 \right. } } \, ,
\end{equation}
$\rho$ is the local baryon density, $U^\rho$ is of the form (\ref{U=}) and the average $\langle \cdot \rangle$ is
over the local particle distribution.  The parameters in (\ref{v=}) and (\ref{U=}) have been adjusted to ground-state nuclear
properties and tested in comparing data on collective flow to transport-model predictions \cite{dan00}.
The parameter set for $a$, $b$, $c$, $\nu$ and $\lambda$ is employed and yields $p_F/v_F^* = 0.70 \, m$
at Fermi momentum and a soft equation of state \cite{dan00}.

For solving the Boltzmann equation set, the test-particle global-ensemble method is employed.  In the test-particle
representation, with $\mathcal{N}$ test particles per genuine particle, the phase-space distribution for $Z$ protons
is, e.g.:
\begin{equation}
f(\overrightarrow{r},\overrightarrow{p},t) = \frac{(2 \pi)^3 }{ {\mathcal N}}\sum_{i=1}^{Z {\mathcal N}}
\delta(\overrightarrow{r}-\overrightarrow{r}_{i}) \, \delta(\overrightarrow{p}-\overrightarrow{p}_{i})\quad ,
\end{equation}
where the sum is over proton quasiparticles.  Because of the use of a global ensemble, effectively each of the
quasiparticles can collide with the quasiparticles in its vicinity at a reduced cross-section $\sigma/\mathcal{N}$.
In detail, the algorithm for quasiparticle collisions is described in \cite{dan}.  Reasoning that the size of
in-medium nucleon-nucleon cross-sections must be limited by in the excited medium by the interparticle distance, in medium
cross-sections have been adopted and tested in \cite{dan00}, of the form
\begin{equation}
\sigma(\rho)=\rho^{-2/3} \, \mbox{tanh} (  \rho^{2/3} \, \sigma_{free}) \, .
\end{equation}
At low densities, these cross-sections reduce to free-space cross-sections, while at high densities they get
limited by the interparticle distance.

The reconstruction of the transient source in a reaction is done by calculating, in the frame of the overall system, the
mass number $A$,
energy $E$ and momentum $\overrightarrow{P}$
for nucleons that are bound in a local frame.
The source excitation energy $E^*$ is then calculated as
\begin{equation}
E^{*} = \sqrt{ E^2 - \left(c \, \overrightarrow{P} \right)^2} - A \, \left(m_N \, c^2 - \frac{B}{A} \right) \, ,
\end{equation}
where $m_N$ is the rest mass of a nucleon and $B/A \approx 9$~MeV is the mean binding energy
for the respective heavy nuclei,
in the transport-theory initializations.

The employed transport model can describe the production of
deuterons and $A=3$ clusters.
Deuterons are created in
a three-body process, n+n+p $\rightarrow$ d+n or p+n+p $\rightarrow$ d+p.
The created deuterons move according to their Boltzmann equation~(\ref{BUU})
and they can be broken up in collisions.
In the same spirit, $^{3}$He and $^{3}$H are created in four-nucleon processes.

\section{Experimental setup}
\label{experiment}

Experimental data for proton-induced reactions on $^{197}$Au were obtained in
the experiment E900 at the Brookhaven AGS accelerator, using the
Indiana Silicon Sphere (ISiS) detector array \cite{hsi}.
This array contains 162 gas-ion-chamber/silicon/CsI telescopes \cite{kwia}.
The data were collected for four projectile momenta: 6.2, 10.2, 12.8 and 14.6 GeV/c.

The experimentally measured charged-particle spectra were separated
into two parts: the fast and the thermal-like components~\cite{lef1,beau}.
The thermal-like source was reconstructed by subtracting all
first-stage ``fast'' particles (E\(\ >\) 30MeV for Z=1 and E\(\ >\) (9Z+40) MeV for Z $\geq$2) from the
target charge and mass, as described in Ref.\ \cite{lef1}.
All thermal particles were assumed to be
less energetic than the above cuts.  The number of first-stage neutrons
was assumed to follow, on the average, a proportionality to the number of first-stage
protons: $M_{n}^{fast}=1.93 \times M_{p}^{fast}$, and other possibilities \cite{lef3} have been
explored as well.
Full details
of the procedures are provided in \cite{lef3}.

The excitation energy of the thermal source is given by \cite{beau,lef3}:
\begin{equation}
E^{*} = \sum_{i=1}^{M_{c}}K_{i} + M_{n}K_{n} + Q + E_{\gamma},
\end{equation}
where K$_{i}$ is the kinetic energy of the i-th detected charged thermal fragment, M$_{c}$ is
the total multiplicity of all detected thermal fragments, M$_{n}$ is the estimated
mean multiplicity of thermal neutrons, K$_{n}$ is the mean kinetic energy of neutrons
(from the SMM model~\cite{smm}),
Q is the difference in mass
between the final products and a cold initial thermal source,
and E$_{\gamma}$ is the estimated energy of deexcitation through gamma emission.

\section{Effects of the momentum-dependent potential and the in-medium cross-sections}

Figure \ref{evolcomp} shows the time evolution of calculated excitation energy,
mean density, entropy per nucleon and mass loss of the source, for different options of the
transport model: with free cross-sections and momentum-dependent
potential ($\sigma_{free},U^p$), with free cross-sections and only density-dependent
mean-field ($\sigma_{free}$), with in-medium cross-sections and momentum-dependent
potential ($\sigma_{in},U^p$), with in-medium cross-sections and only density-dependent
potential ($\sigma_{in}$), and, finally, with in-medium cross-sections and momentum-dependent
mean-field ($\sigma_{in},U^p$), but no light composite particle (LCP) production.
In comparing the results in the figure, we may note that in-medium cross-sections reduce
the energy deposition in the target.  This is because, for lowered cross-sections,
the medium becomes more transparent
to the energetic cascade initiated by the projectile.  Further, we may note in the figure
that a momentum-dependent potential enhances the energy deposition.  This is likely due to
the fact that the momentum dependence enhances the speed of particles in the medium, increasing
the collision rate.  In combination of the effects, out of the first four cases of calculations,
the energy deposition is strongest for the
($\sigma_{free},U^p$) case and weakest for ($\sigma_{in}$).  As far as the overall magnitude is
concerned, however, the energy deposition is quite similar in the four cases.  When we suppress the
composite particle production, per given mass loss, the system cools off faster.  This is because nucleons,
at a given system temperature,
carry off more kinetic energy per unit mass than do composite particles.  The faster cooling also
results in a lower entropy for the source than in the nucleon-only case.  Notably, at a general level,
one expects a higher entropy generation when the composite production is included, since
more degrees of freedom become available to the system.

Figure \ref{number150} shows next the cumulative numbers of different emitted preequilibrium particles, as
a function of time, at $b=2$ fm.  We employed for the figure a somewhat arbitrary cut in the kinetic energy,
at $E_K$=150 MeV, in order to distinguish the pre-equilibrium $E_K < 150$ MeV) from prompt particles ($E_K > 150$ MeV).
At later times in Fig.\ \ref{number150}, we can see that the emission of the preequilibrium particles is enhanced when the free cross-sections are used.  Otherwise, up to $\sim 40$ fm/c the emission is rather independent of the cross-sections.  This is because the particles which come out first are fast and stem from high-energy
processes taking place at low cross-sections, little modified by the medium.
The momentum-dependence of the mean-field speeds up the passage of particles through the medium.
On one hand, this dependence enhances early particle emission, as apparent in the figure.  On the other hand,
the momentum dependence somewhat increases the production of composite particles in general.
Specifically, the production requires several particles present within the same location and, thus, it is enhanced
at higher densities in a reaction.  Higher speeds of energetic nucleons, relative to the medium,
permit for more production of the composites, before the overall medium density decreases.
In comparing the temporal evolution of the emission of different particles relative to each other in Fig.~\ref{number150},
we can see that the times when these emissions become significant are different for different particles,
with the emission of protons setting in at $\sim 15$ fm/c, and the emission of deuterons and tritons
setting in, respectively, at $\sim 22$ and $\sim 27$ fm/c.  These differences can be understood in terms
of the difference in speed of the different particles, at comparable energies or momenta,
particularly with the composites needing to
move out from the higher-density interior of the source.

Figure \ref{ekmeanpre} shows next the mean kinetic energy of
protons, with $E_K < 150$~MeV, as a function of the proton emission time, at different impact parameters.
The particular results refer to the ($\sigma_{in},U^p$) case of the calculations.  At short times,
when the emission rates are low, the averages fluctuate.
If we adopt an energy cut-off of 30~MeV \cite{morley}, to separate the non-equilibrium from
equilibrium emission, we arrive at thermalization times $\tau \gtrsim 40$~fm/c.  The thermalization times
interestingly appear to be nearly independent of the impact parameter.

\section{Comparison to experimental data}

For the reaction p+$^{197}$Au, isotope-resolved experimental data on the light charged particles exist \cite{beau} for
four incident momenta of 6.2, 10.2, 10.8 and 14.6 GeV/c.  It is therefore of interest to compare the data from this simple
p+A reaction, to the transport-model
predictions on the early-stage cluster formation.
The light charged particle ratios, d/p, t/p and $^3$He/p, from the ($\sigma_{in},U^p$) simulations of the 14.6 GeV/c
reaction, are compared in Fig.~\ref{ratio14im} to the data, as a function of the excitation energy 
and at different times in the reaction,
both before and after an application of
the filter simulating the effects of experimental setup and procedures.
The unfiltered ratios are shown in the left-hand and the filtered in the right-hand panels.
The theoretical excitation energies represent averages for specific impact parameters.
The filter is based on the geometrical arrangement of detectors
and on the kinetic energy acceptance for non-thermal charged particles.
Specifically, to simulate the experimental
selection of the particles emitted before thermalization, the energy
acceptance ranges from 30 MeV for Z=1 and 58 MeV for Z=2 up to the
upper detection limit of the detectors (E$_k$\(\leq \)350MeV
for p, E$_k$\(\leq \)136MeV for d and E$_k$\(\leq \)162MeV for t).
The main effect of the geometrical part of the filter is to eliminate
the unmeasured forward-focused (below 14 degrees) energetic ``prompt'' protons
emitted early in the collision. The energy cuts eliminate very high energy
protons, but also slow particles that tend to be emitted later in the simulations (after 40-45
fm/c).
The comparison between the theory and data is done for the
energetic, dynamical, rather than thermal, particles.

As can be seen in Fig.~\ref{ratio14im}, the filtered-particle ratios generally stabilize
past 50 fm/c.  If we make a comparison at 65 fm/c, we find a semiquantitative agreement
between the theory and data for the t/p and d/p ratios and a bit poorer agreement
for the $^3$He/p ratios.  The $^3$He/p ratios remain below the data no matter what
later times in the simulations are used for the comparison.  Overall, we find that
the experimental ratios of Z=1 isotopes as a function of excitation energy can be reproduced, with some success,
at late times in the filtered transport-simulations.

The experimentally reconstructed average source charge $Z_{src}$ and the
mass loss $\Delta A_{src}$ for the source are next represented in Fig.~\ref{zvse}, 
as a function of the reconstructed excitation energy,
for the 6.2 GeV/c (left panels) and
14.6 GeV/c (right panels) reactions.
The shaded area in the 6.2 GeV/c mass-loss panel illustrates uncertainties that are reached when following different hypotheses\cite{lef3}
in estimating the first-stage neutron multiplicity; the circles represent results
obtained with the relation indicated in Sec.\ III.
Overlaid over the results from measurements are the theoretical results for the source, stemming
from the ($\sigma_{in},U^p$) simulations,
at the different times $t$ in a reaction.  The results for $t=65$~fm/c agree reasonably well with those
obtained from data when following the relation of Sec.\ III.

Figures \ref{compz} and \ref{comprat} show next, for the p+Au reaction at 14.6 GeV/c,
similar comparisons to those in Figs.\ \ref{zvse} and \ref{ratio14im}, but now confronting
the different assumptions on nucleon-nucleon cross-sections and mean-field potentials.
We observe, as in the context of Fig.\ \ref{evolcomp}, that the effect of
the momentum-dependence in the mean-field is to allow the source to reach
higher excitation energy (by $\sim$150-200 MeV), better conforming with data.
As to in-medium cross-sections, we find in Fig.\ \ref{zvse} that they have little impact on the relations
between the charge and mass of the source and the excitation energy.  On the other hand,
only the in-medium cross-sections allow the simulations to reproduce the particle ratios
consistently with also reproducing the source characteristics.  Overall the best agreement of 
simulations with data, for a late time
in a reaction, is reached for the ($\sigma_{in},U^p$) version of the simulations.

If the thermalization time is defined as that time for which the simulation best reproduces
the experimentally reconstructed source, perceived to be the basis of thermal emission, then
the time, judging from the ($\sigma_{in},U^p$) results in Figs.~\ref{ratio14im}-\ref{comprat}, is close to
65 fm/c, both at 6.2 and at 14.6 GeV/c.  This time is somewhat higher than that first deduced on
the basis of Fig.~\ref{ekmeanpre} ($\gtrsim 40$~fm/c) but the former time principally represents
a lower bound on the thermalization.  Obviously, one also needs to remember that, in general, no perfect match can be reached between
conclusions reached in the experiment from spectra and in the simulations from a temporal progress.
Overall, in the simulations we see that the bulk of the energy deposition
process (associated with a saturation of the entropy) is over after about 25 fm/c (see Fig.~\ref{evolcomp}).
This leaves $\sim 40$ fm/c for redistributing the energy across the nucleus, i.e.\ to thermalize.
In detail, however, some emissions might take place early, on which, could be classified as thermal
on the basis of spectra; conversely some emissions might take place later that could be classified as
first-stage spectra.

The effect of light composite emission on the relation between different source characteristics at
different times is next explored in Fig.~\ref{zvsenc}, which shows the source charge (Z$_{src}$) and the mass loss ($\Delta
A_{src}$) as functions of E*, for BUU without the production of light clusters at 14.6 GeV/c.
It is seen that the deduced thermalization time would have been earlier than
for the dynamics with light-particle, with best agreement with data achieved at $t \sim 50$~fm/c.
However, the maximal excitation energies of $\sim 800$ MeV turn out to be well below that
observed experimentally compared to the simulations with clusters, which extends to $\sim 1400$ MeV.

In comparing to data, it may be interesting to construct a distribution of the excitation energies
for the simulations.  The simulations, equally spaced in the impact parameter, are weighted with
the factor
\begin{equation}
\sigma (b) = \pi (b_{max}^2-b_{min}^2) \, ,
\end{equation}
where $b_{max}$ and $b_{min}$ are the upper and lower ends of the $b$-interval that the simulation
represents.  Furthermore, for each impact parameter, we assume the $E^*$-distribution
to be of a gaussian form, with an assumed value for the full width at half maximum (FWHM).
Figure \ref{disring} next compares the constructed distributions, for different versions of the
simulations and different FWHM, to the distributions from 14.6 GeV/c data.  The distributions are
normalized at $E^* = 400$~MeV.  The left-hand panels display distributions constructed with a FWHM
of 250 MeV and the right-hand panels - with a FWHM
of 400 MeV.  The higher value of FWHM is favored by comparisons and the ($\sigma_{free},U^p$)
simulations yield predictions inferior to other simulations.

Overall, we find that the comparison between data and calculations favors the ($\sigma_{in},U^p$)
version of the simulations and suggests the thermalization time of the order of 65 fm/c at both
incident momenta.

\section{Transport model predictions}

The predicted evolutions of the average excitation energy and of the average density
of the source formed in 14.6 GeV/c p+\(^{197} \)Au reactions at different
impact parameters, from ($\sigma_{in},U^p$) transport calculations, are displayed
in Fig.~\ref{e_rho_time}.  At all impact parameters $b$, the deposited energy
increases strongly during the first few fm/c, as the projectile crosses more of the
target.  The excitation energy reaches a maximum when the projectile leaves the
target.  The respective times for the maxima then change with the impact parameter,
from 18 fm/c for the most central collisions to 10-12 fm/c for more peripheral.
With the maximum reached earlier in the peripheral collisions, for the shorter
proton trajectories inside the target, the energy deposition is lower in the peripheral
collisions.

The average source density in Fig.~\ref{e_rho_time} appears nearly constant
during the first 10 fm/c of the collision, independent of the impact parameter.
However, this density decreases rapidly thereafter as the energy
dissipation increases and more ``prompt" nucleons get ejected.
The nucleon knock-out process, the so-called ``swiss-cheese'' effect, appears actually to be
the driving factor in depleting the nuclear density.  The progress of that process
depends on the impact parameter and its effects mimic those of an expansion.  To illustrate the process, Fig.~\ref{xy} shows
the evolution of the density in p+Au reaction at 14.6 GeV/c and $b=2$~fm.
Unlike in Wang {\em et al.}\ \cite{wang}, we do not quite see a cavitation in this reaction,
but, nonetheless, we can see quite clearly a local depletion of density along
the trajectory of the projectile, as prompt nucleons are ejected on a
very short time scale.  The depletion becomes particularly obvious at around 20 fm/c
right after the deposited energy maximizes, as further exhibited in Fig.~\ref{evol}.
It is not until
20-30 fm/c later that the system relaxes and that the nuclear density
smoothes out.  It is also after the first 20 fm/c that the most
significant mass loss occurs, through prompt emission processes,
cf.\ the bottom panel of Fig.~\ref{evol}.

Figure~\ref{evol} generally explores the effect of entrance channel beam momentum
on the source evolution, displaying the excitation energy, average density,
entropy per nucleon and mass as a function of time, for the $b=2$~fm p+Au reactions
at 6.2, 10.2, 12.8 and 14.6 GeV/c.
It is seen that the deposition of energy, the generated
entropy per nucleon and the mass loss for the source increase with the beam
momentum, while the mean density decreases.
After 15-20 fm/c, the average density of the system decreases rapidly
to reach 1/3 to 1/4 of the normal density by 40 fm/c,
after which the
entropy per nucleon becomes relatively stable for all systems.
It is at that point that the chaotic regime appears to be established and
the deposited energy can be attributed to the thermal source.

Increasing mass loss, due to an ejection of prompt nucleons and then the preequilibrium
emission, with increase of the incident momentum, is consistent with what has been observed
experimentally \cite{beau}.  With increase in the excitation energy and the number of
collisions and produced light fragments, the entropy per nucleon increases
with the incident momentum.  At any momentum, the entropy per nucleon stabilizes at about the same time as thermalization, which also coincides with a decrease in nuclear density.  On one hand,
when the system is close to thermalization, the gains in entropy during evolution become
generally moderate.  On the other hand, when density drops, the collision rate decreases
and entropy generation slows down.  Because of evaporation, in fact, the entropy per nucleon
of the source itself may begin to drop.

It may be further of interest to explore the evolution of different
isotope ratios with time.  Figure~\ref{ratiotime} shows the unfiltered ratios
of d/p, t/p, t/$^{3}$He and $^{3}$He/p, as a function of time, from the ($\sigma_{in},U^p$)
simulations at 14.6 GeV/c and different impact parameters.
For all
ratios relative to protons, there is a common ordering: the smaller the
impact parameter, the higher the ratio. This can be linked to larger nucleon
yields, emitted over a shorter time in the more central reactions. To form a cluster, several nucleons need to move simultaneously out of the system. The t/$^{3}$He ratio, lower in more central than in peripheral collisions is possibly due
to larger neutron relative to proton loss in central collisions, driving
the system towards symmetry, cf. Fig.\ \ref{compz}.

Finally, for the optimal thermalization times, when experimental relations
between the source charge and mass and excitation energy are reproduced, we examine
the thermalized source characteristics at the different incident momenta.
Figure \ref{svse} shows the dependence of the average density and entropy per
nucleon from the ($\sigma_{in},U^p$) simulations, as functions of excitation energy in the different reactions.
Each
point in the construction of the plots corresponds to a specific impact parameter,
between 0.4 fm to 8.0 fm,
for a given incident momentum.  It is apparent that for each of the incident momenta,
the density and entropy have a fairly smooth dependence on the excitation energy.
Though, at a given
excitation energy, there is a small decrease in the average density and a slight increase
in the entropy per nucleon with a rise in the incident momentum, primarily
the dependencies on the excitation energy are fairly universal with the momentum.  There is some increase
in the range of excitation energy between 6 and 10 GeV/c and less afterwards.  Notably,
at the excitation energy of 1000 MeV (corresponding to 5-6 MeV/nucleon), the source becomes
quite dilute, reaching densities between 1/4 and 1/3 of normal density. The system becomes even
more dilute at higher excitation.

\section{Conclusions}
Proton-induced reactions provide a baseline
comparison for any models constructed to explain the A+A reactions.
In this paper, we have confronted a comprehensive data set for the p+$^{196}$Au reaction between 6.2 GeV/c and 14.6
GeV/c with a transport model including light-cluster production.  We have concentrated on the
characteristics of the thermal source in the reactions, its mass, charge and excitation energy,
and on the normalized yields for pre-equilibrium LCPs.
We found that the cluster formation within the model, during the pre-equilibrium stage, enhances the source excitation energy
compared to the situation without clusters. Judging from the entropy, the majority of dissipation leading to the formation of a thermalized source takes place over the first 25-30 fm/c of
a reaction. However, an additional 35-45 fm/c is required for the pre-equilibrium particles to leave the source
and for the source to acquire characteristics that can be identified experimentally
for the thermal emission.

Filtered yields for pre-equilibrium LCPs have been analyzed as a function of excitation
energy.  We found that the use of in-medium scattering cross-sections and momentum-dependent potentials considerably improved the agreement with data, compared to earlier, more simplified versions of the calculations.
Nonetheless, we could not obtain a simultaneous complete agreement with the measured $^3$He yield.
If the model was extended to the alpha particles, it would be certainly interesting to look in the future at the pre-equilibrium emission
of alphas and the effect of that emission on energy dissipation.  It is entirely possible that the
production of alphas impacts the production of $^3$He.

An optimal characteristic time of $\sim 65$~fm/c was extracted for the p+$^{197}$Au reactions, for
an agreement between the measured and calculated source characteristics and pre-equilibrium particle-ratios.
The optimal time was found to be approximately independent of the impact
parameter and of the incident momentum.  The transport model predicts highly excited sources, with excitation
energies up to 5-6~AMeV, at low densities, down to 1/4-1/3 of normal density,
i.e.\ close to the typical multifragmentation conditions.

The target nuclei undergo a density depletion from early on in the reaction,
as nucleons and light clusters are ejected along
the projectile trajectory.  The ejection itself reduces the density, which
is followed by a readjustment in the target density and momentum distributions.
We do not, however, observe the development
of a spherical cavity, nor development of strong compression waves.  Rather,
the system arrives at the thermalized state at $\sim65$~fm/c, at densities close to the critical density,
in the course of a gradual healing of the cheese-effect due to the prompt cascade,
involving the pre-equilibrium particle emission.  The features of the system then
lend support to the analysis methods of nuclear
fragmentation relying on the concepts of equilibrium within the phase transition region
as applied to the ISiS data~\cite{Berk,Ell1}.

\vspace{0.5cm}
{\bf \centerline{ACKNOWLEDGEMENTS}}

This work was supported by the Natural Sciences and Engineering Research
Council of Canada, the Fonds \`a la Recherche et l'Aide au Chercheurs
(FCAR) Qu\'ebec, by the U.S. Department of Energy, the National
Science Foundation and the Robert A. Welch Foundation.

\newpage

\begin{figure}
\caption{\label{evolcomp}
Excitation energy, mean density, entropy per nucleon and mass of the
source as a function of time from the transport-model simulations of the 14.6 GeV/c p+$^{197}$Au reaction at
$b=2$~fm.  The different lines and symbols represent results for different versions
of the simulations, as indicated.}
\end{figure}

\begin{figure}
\caption{\label{number150}
Number of different emitted charged fragments, with $E_k < 150 $ MeV,
as a function of time in the transport-model simulations of the 14.6 GeV/c p+$^{197}$Au reaction at
$b=2$~fm. The different lines and symbols represent results for different versions
of the simulations, as indicated.}
\end{figure}

\begin{figure}
\caption{\label{ekmeanpre} Mean kinetic energies of emitted pre-equilibrium protons versus emission time
for ($\sigma_{in},U^p$) simulations of the 14.6 GeV/c p+Au reaction at different impact parameters.
The dashed line represents the experimental low-energy acceptance cut.}
\end{figure}

\begin{figure}
\caption{\label{ratio14im} Light charged particle yield ratios as a function of excitation energy in the
p+Au reaction at 14.6 GeV/c.  The top, center and bottom panels show, respectively, the d/p, t/p
and $^3$He/p ratios.  The circles represent data and the lines represent results of ($\sigma_{in},U^p$)
calculations at different times in the reaction.
The left and right panels show, respectively, results of calculations before and after filter application. }
\end{figure}

\begin{figure}
\caption{\label{zvse}
Source charge number (top panels) and mass loss (bottom panels) as a function
of excitation energy in 6.2 GeV/c (left panels) and 14.6 GeV/c (right panels) p+\(^{197} \)Au  reaction.
Circles represent results from data obtained following the procedure from Sec.\ \protect\ref{experiment}.
Shaded area in the left bottom panel illustrates uncertainties in the experimental extraction
associated with the assumptions on the initial fast neutrons \protect\cite{lef3}.  The lines represent
results of the ($\sigma_{in},U^p$) transport-model simulations at different times in a reaction. }
\end{figure}

\begin{figure}
\caption{\label{compz}
Average source charge number $Z_{src}$ and mass number loss $\Delta A_{src}$ as a function of the excitation
energy $E^*$ for the reaction p+$^{197}$Au at 14.6 GeV/c.  The circles represent data.  The lines represent
different versions of transport calculations, at different instances in the reaction. }
\end{figure}

\begin{figure}
\caption{\label{comprat}
Normalized yields of light charged particles as a function of the source excitation energy
the p+Au reaction at 14.6 GeV/c.  Circles represent data and lines represent filtered ratios
from
different versions of transport calculations, at different instances in the reaction. }
\end{figure}

\begin{figure}
\caption{\label{zvsenc}
Average source charge number $Z_{src}$ and mass number loss $\Delta A_{src}$ as a function of the excitation
energy $E^*$ for the reaction p+$^{197}$Au at 14.6 GeV/c.  The circles represent data.
The lines represent the results of ($\sigma_{in},U^p$) calculations without light-composite
production, at different instances in the reaction. }
\end{figure}

\begin{figure}
\caption{\label{disring}
Source distribution in the excitation energy,
normalized to 1 at $E^*$ = 400 MeV, for the p+$^{197}$Au reaction at 14.6 GeV/c.  The histogram
represents data.  The different lines represent different versions of the calculations at different times
in the reaction, assuming different widths FWHM for the excitation energy distributions at individual
impact parameters. }
\end{figure}

\begin{figure}
\caption{\label{e_rho_time}
Source excitation energy (top) and average density (bottom) as a function of both impact parameter
and time, from the ($\sigma_{in},U^p$) simulations of the p+$^{197}$Au reaction at 14.6 GeV/c.
}
\end{figure}

\begin{figure}
\caption{\label{xy}
Nuclear density within the reaction plane XZ, where Z is along the beam axis, at different times
in the p+$^{197}$Au reaction at 14.6 GeV/c and $b=2$~fm, from the ($\sigma_{in},U^p$) transport simulations.
 }
\end{figure}

\begin{figure}
\caption{\label{evol} Excitation energy of the source per nucleon, mean density,
entropy per nucleon and mass number as a function of time, in p+Au reaction at $b=2$~fm and different incident
momenta, from the ($\sigma_{in},U^p$) transport simulations.
 }
\end{figure}

\begin{figure}
\caption{\label{ratiotime} Evolution of the unfiltered light charged particle ratios with time,
at different impact parameters in the p+$^{197}$Au reaction
at 14.6 GeV/c, from the ($\sigma_{in},U^p$) transport simulations.
}
\end{figure}

\begin{figure}
\caption{\label{svse}  Mean density (top) and entropy per nucleon (bottom) of the
source as a function of its excitation energy, for four different
beam momenta in the p+$^{197}$Au reaction, from the ($\sigma_{in},U^p$) transport simulations,
at the time of an agreement with data on the source mass and charge and on the pre-equilibrium particle ratios,
cf.\ Figs.\ \ref{ratio14im} and \ref{zvse}.}
\end{figure}

\newpage

\begin{figure}
\centerline{\psfig{figure=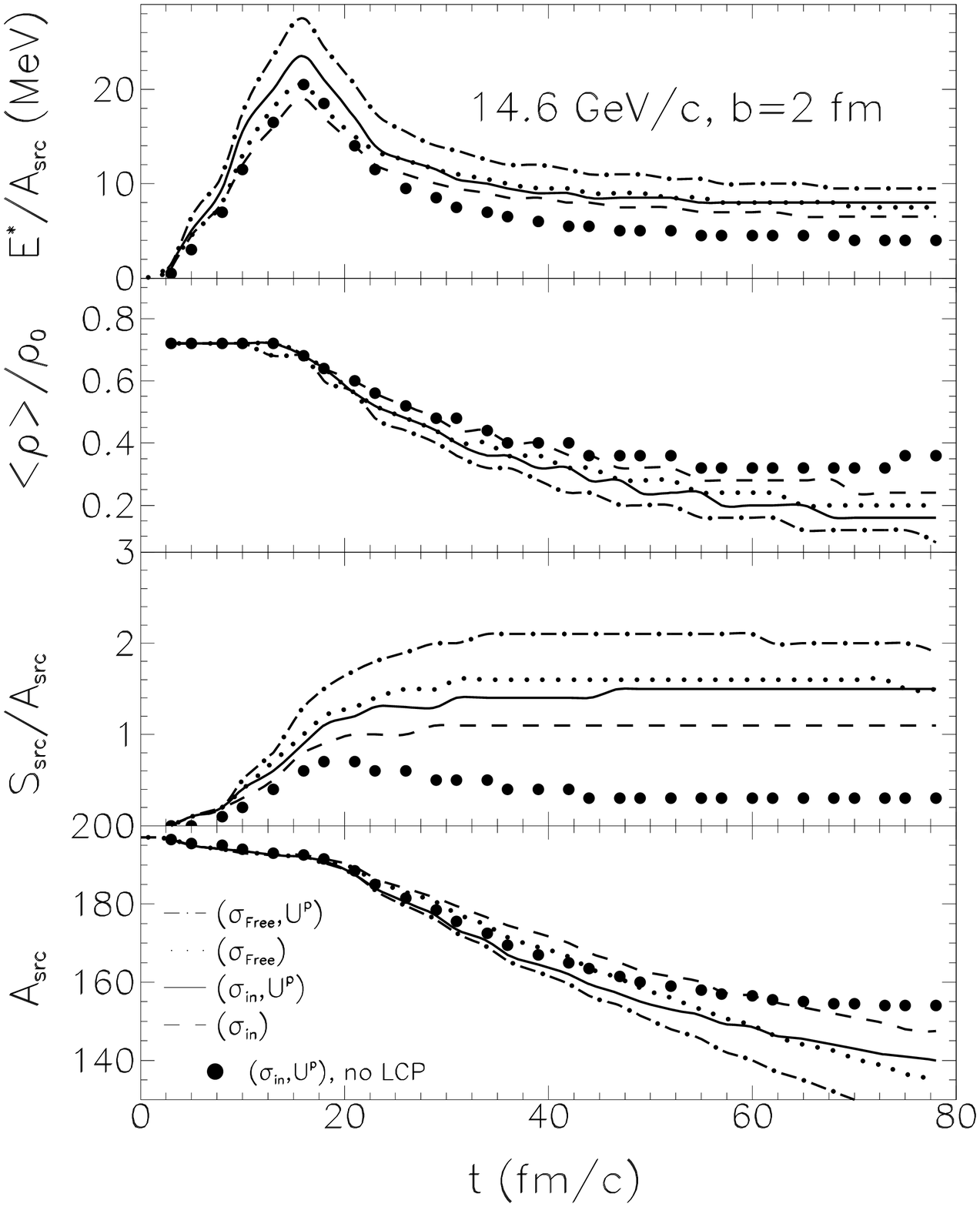,width=5.5in}}

\center{FIG. 1}
\end{figure}

\newpage
\begin{figure}
\centerline{\psfig{figure=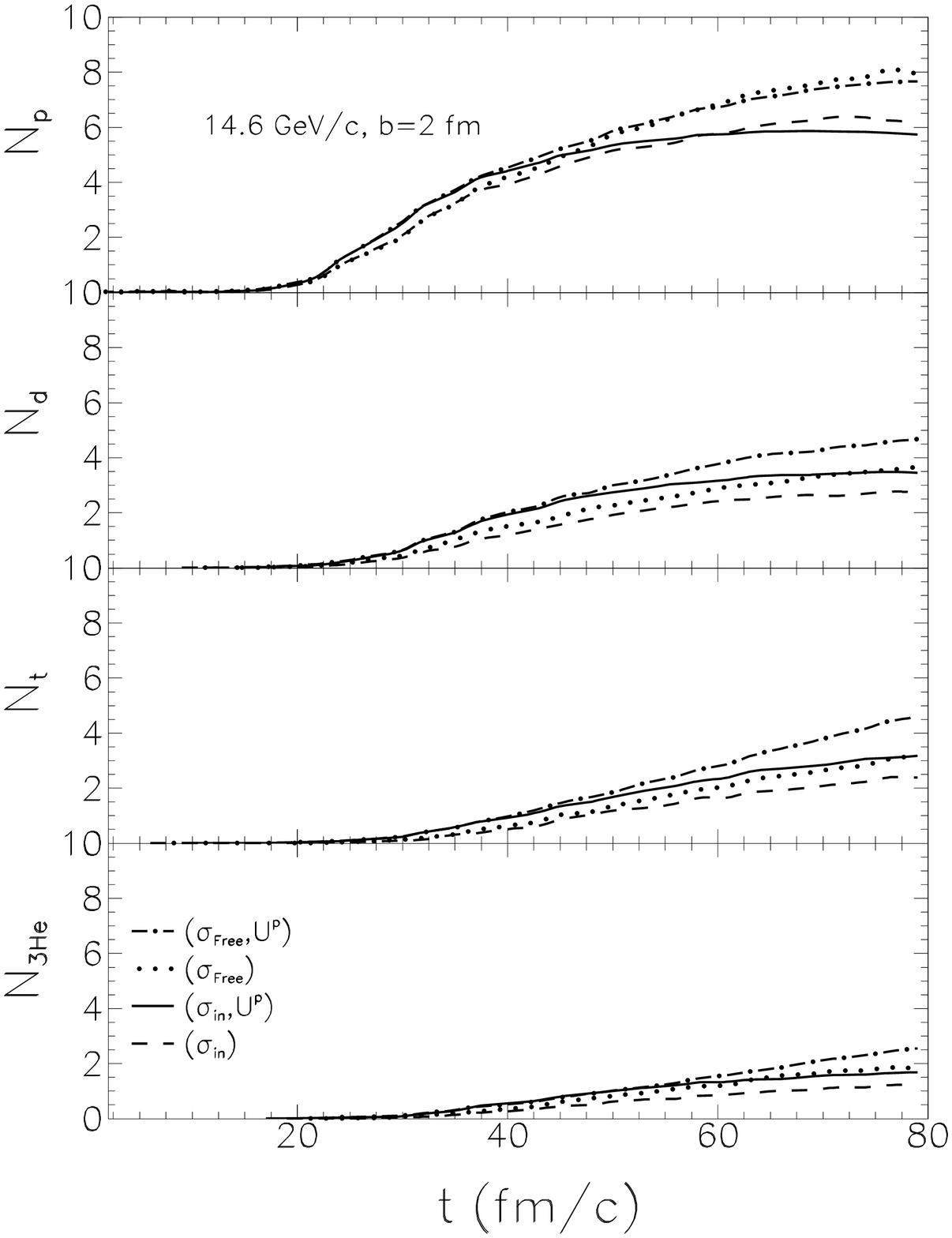,width=5.5in}}

\center{FIG. 2}
\end{figure}

\newpage
\begin{figure}
\centerline{\psfig{figure=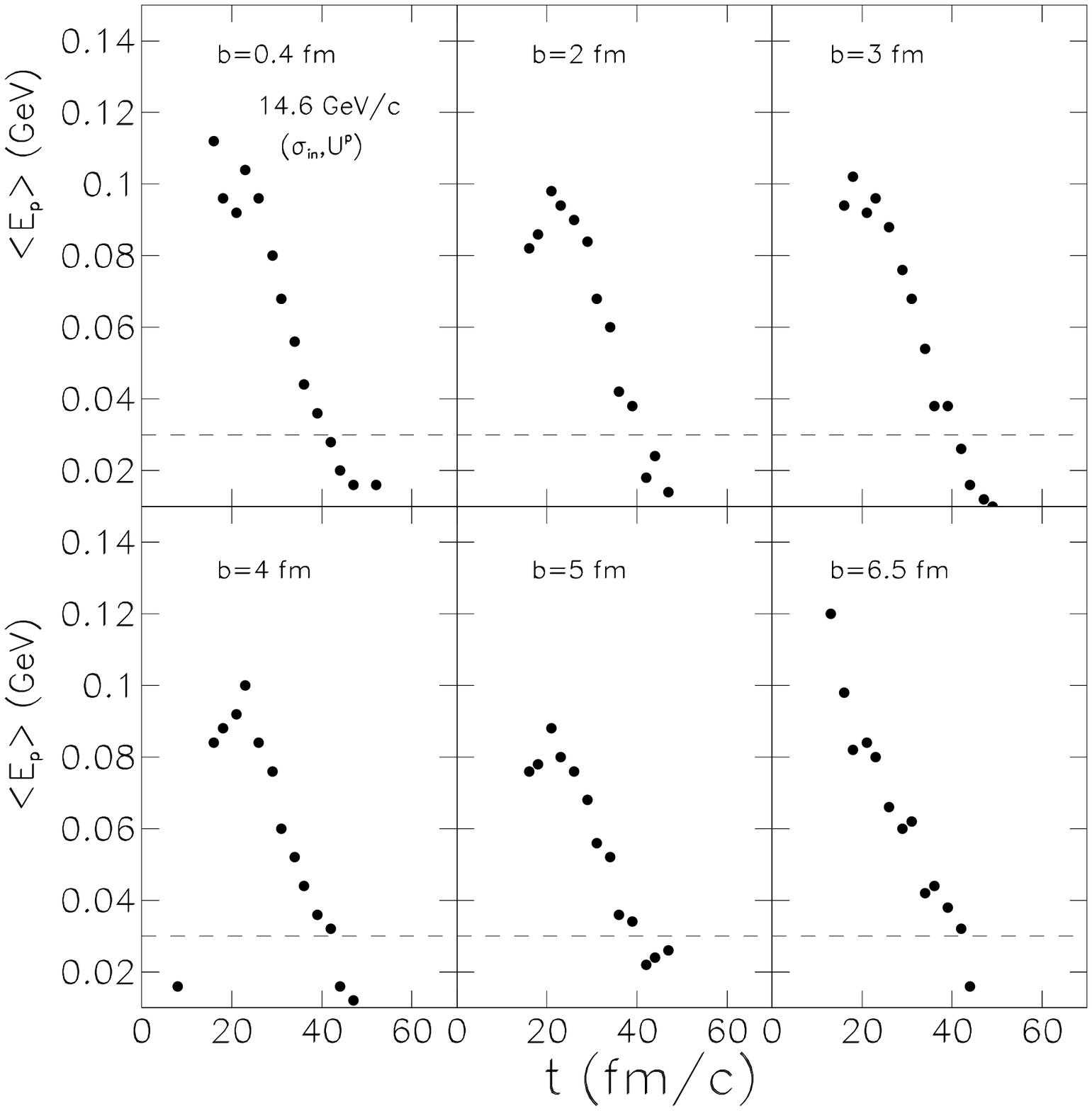,width=5.5in}}

\center{FIG. 3}
\end{figure}

\newpage
\begin{figure}
\centerline{\psfig{figure=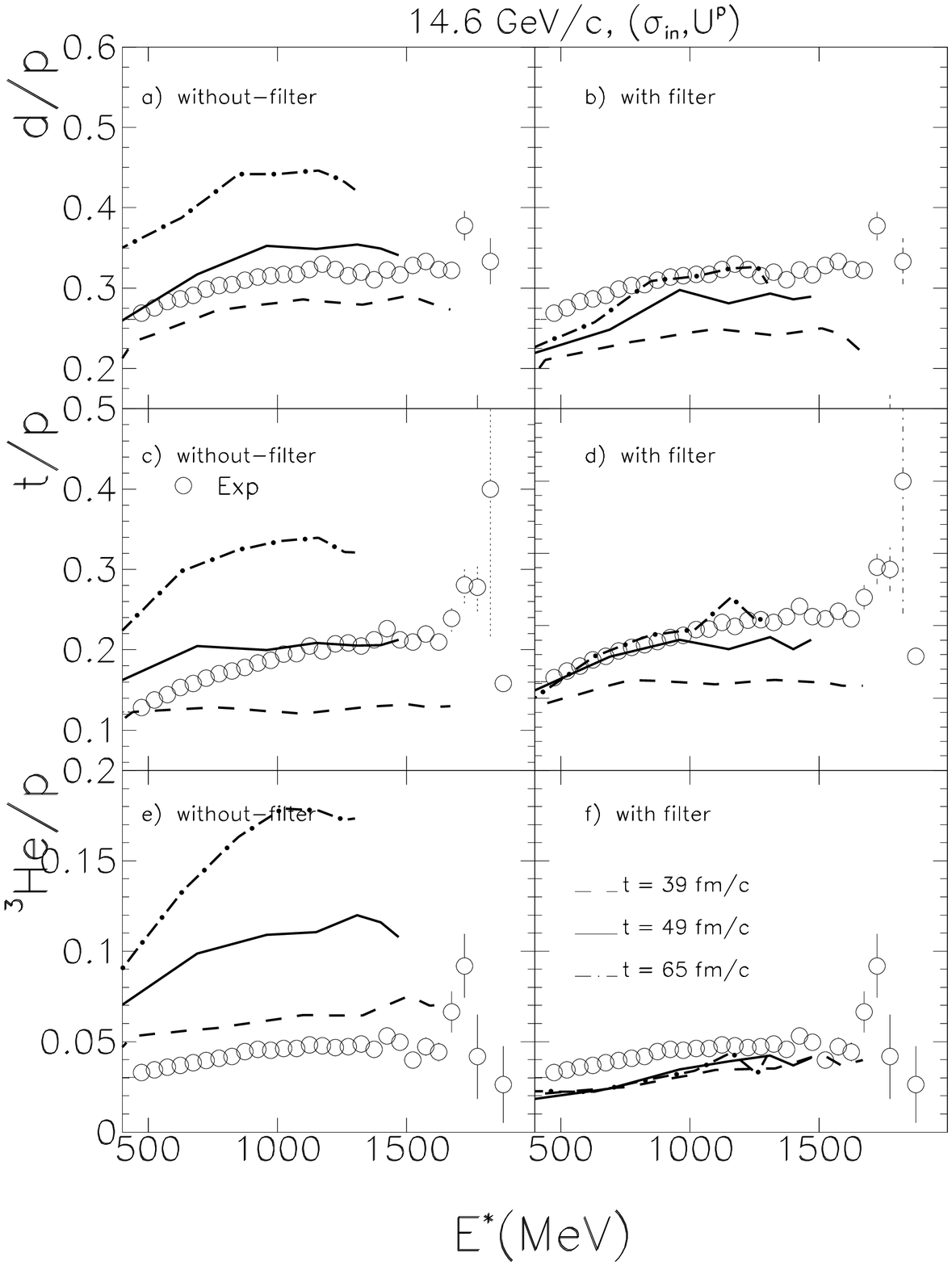,width=5.5in}}

\center{FIG. 4}
\end{figure}

\newpage
\begin{figure}
\centerline{\psfig{figure=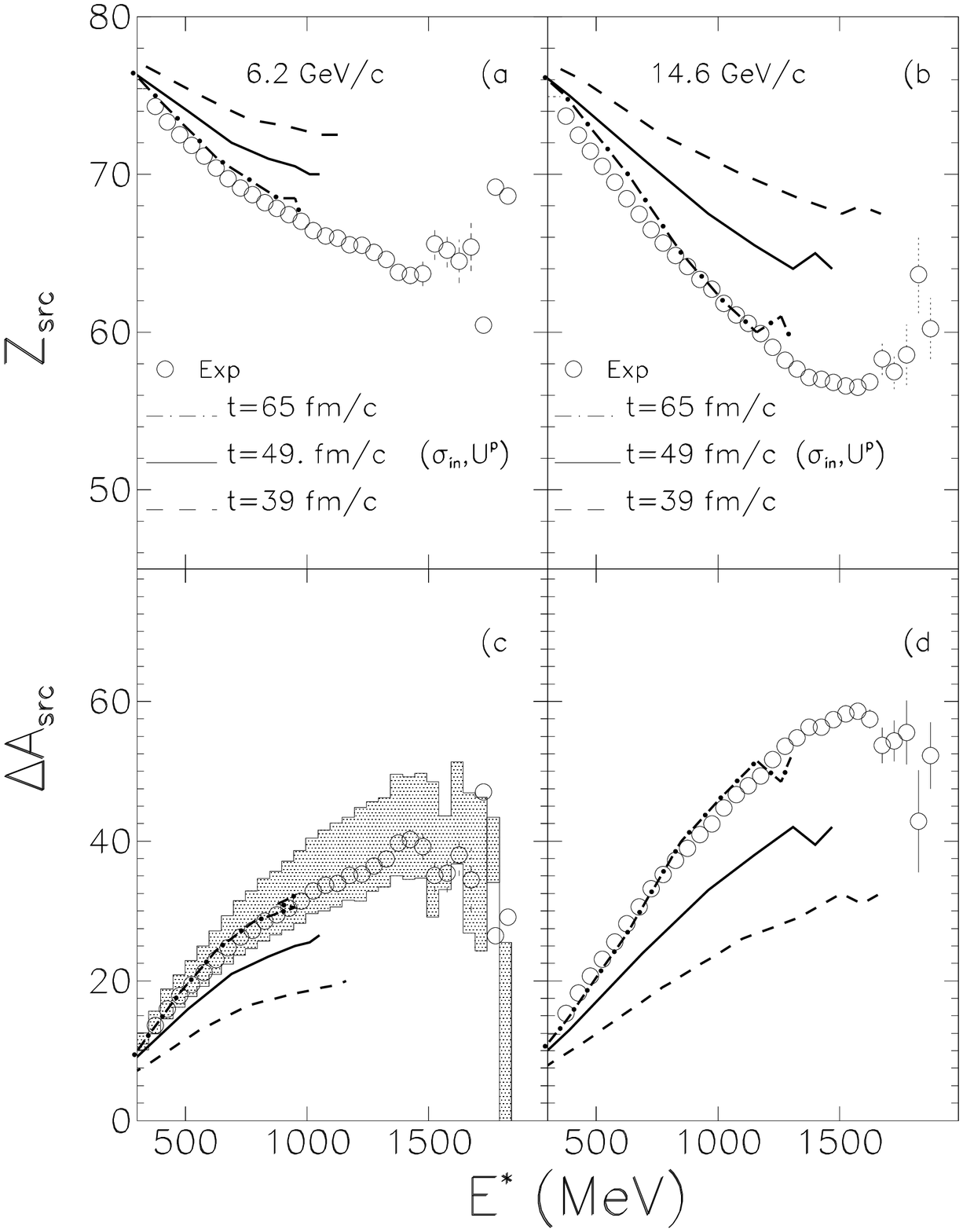,width=5.5in}}

\center{FIG. 5}
\end{figure}

\newpage
\begin{figure}
\centerline{\psfig{figure=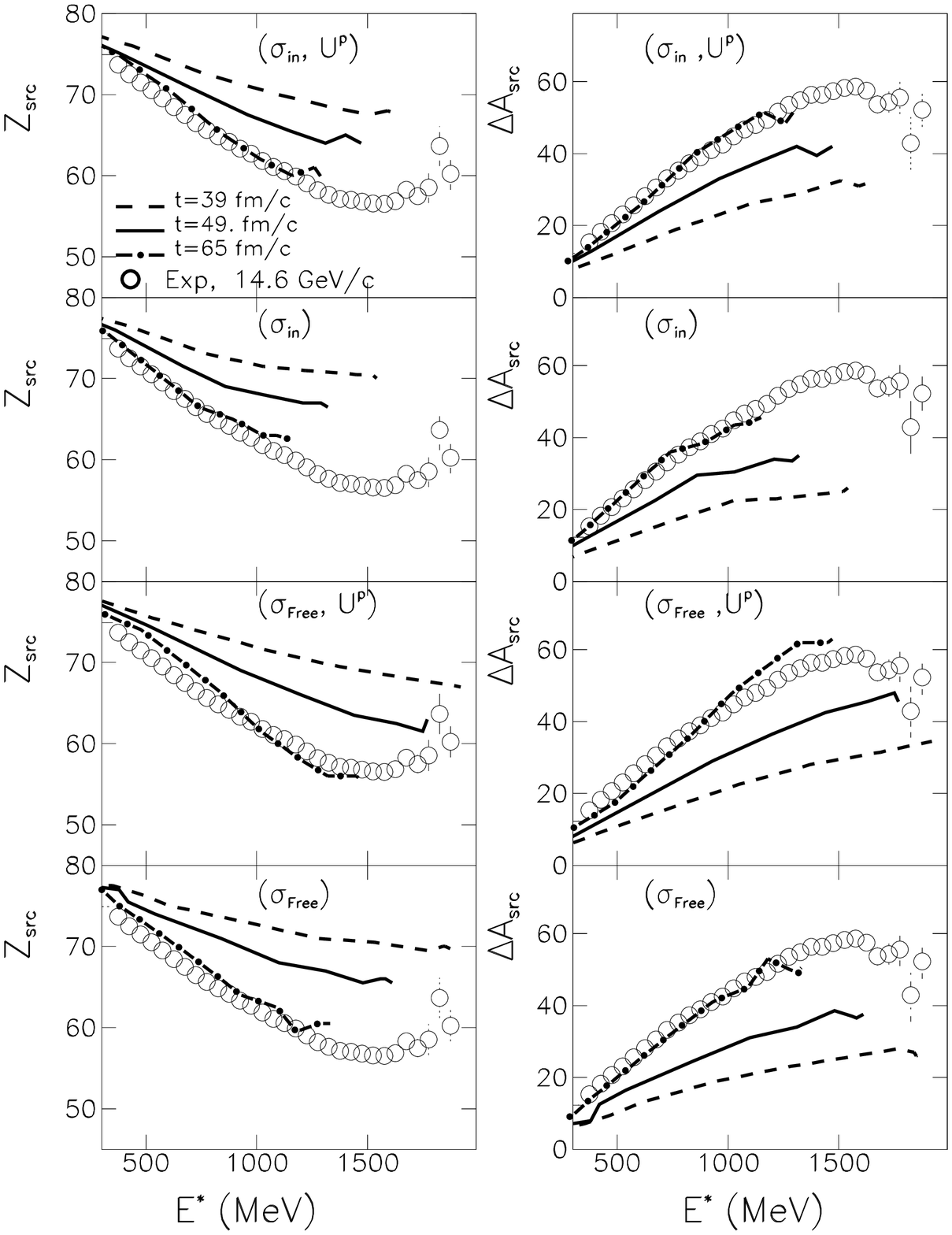,width=5.5in}}
\center{FIG. 6}
\end{figure}

\newpage
\begin{figure}
\centerline{\psfig{figure=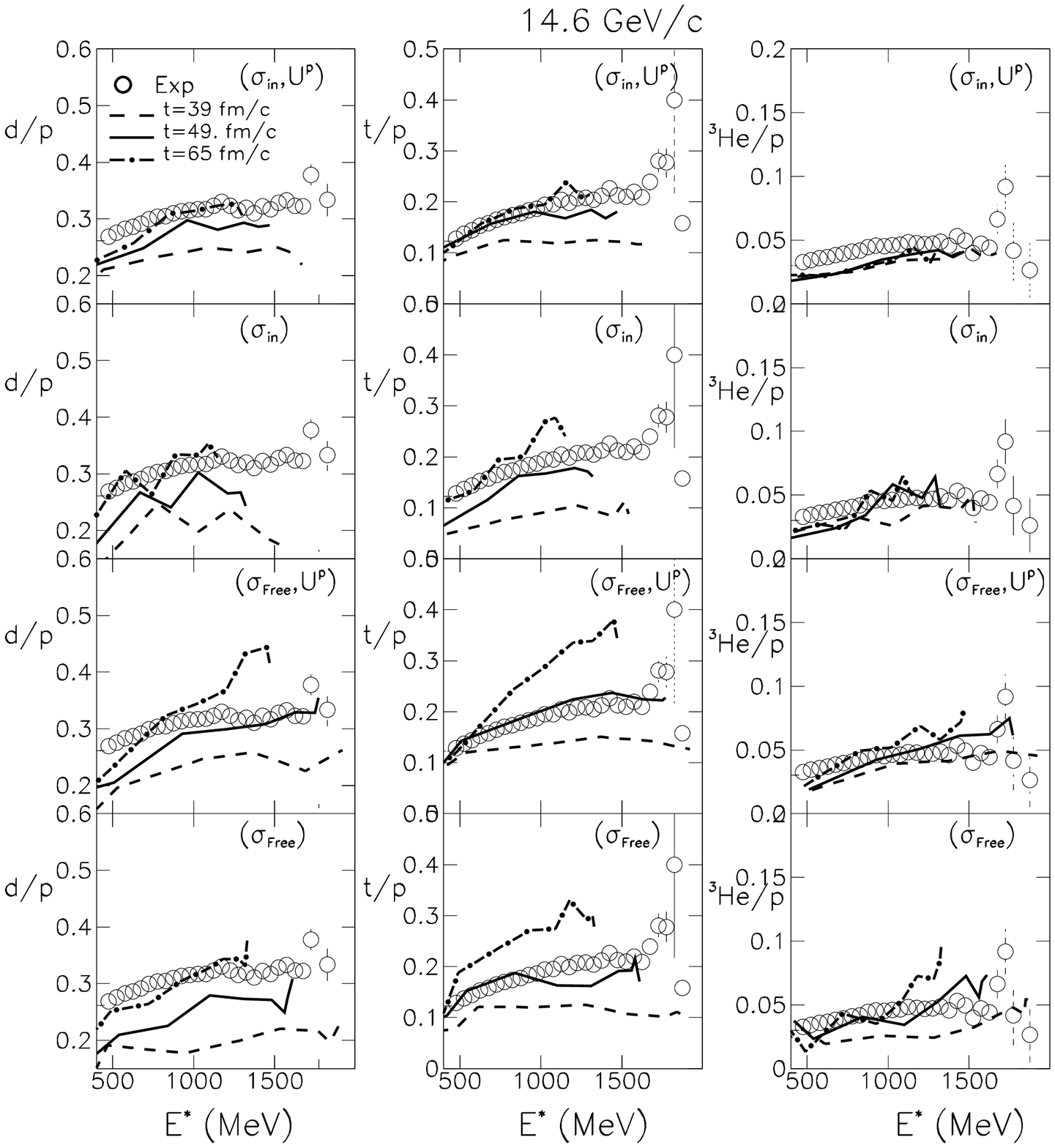,width=5.5in}}
\center{FIG. 7}
\end{figure}

\newpage
\begin{figure}
\centerline{\psfig{figure=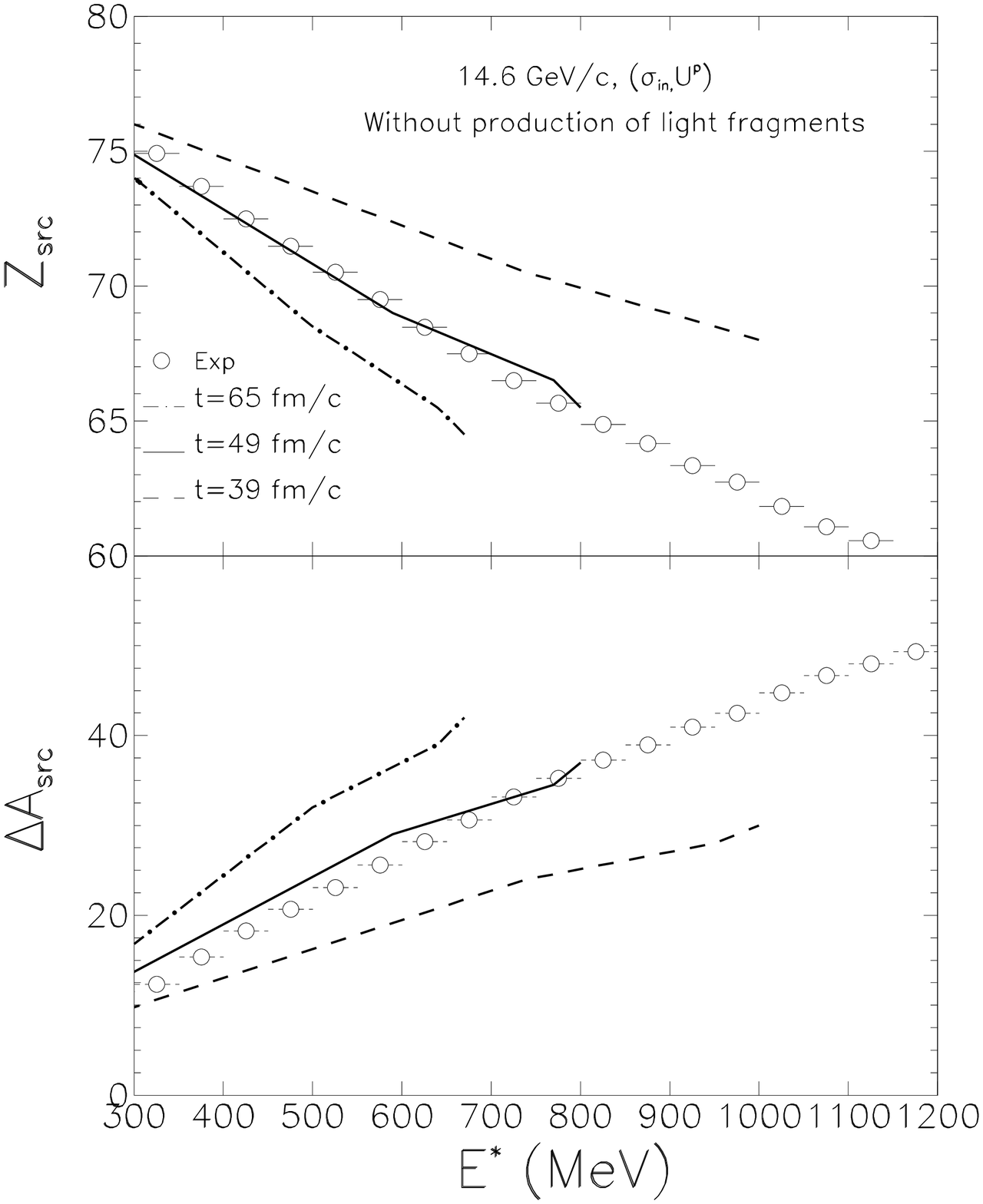,width=5.5in}}
\center{FIG. 8}
\end{figure}

\newpage
\begin{figure}
\centerline{\psfig{figure=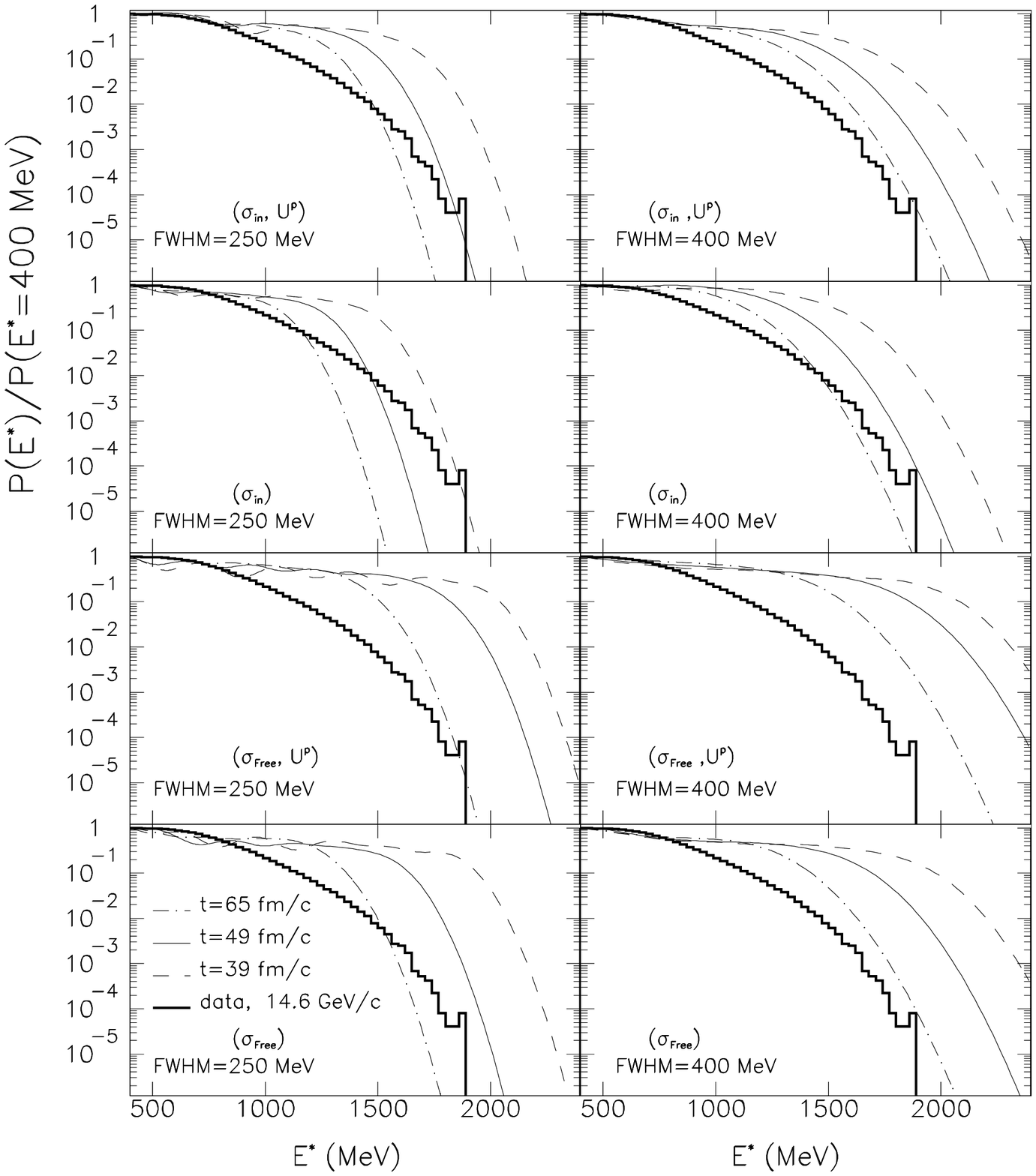,width=5.5in}}
\center{FIG. 9}
\end{figure}

\begin{figure}
\centerline{\psfig{figure=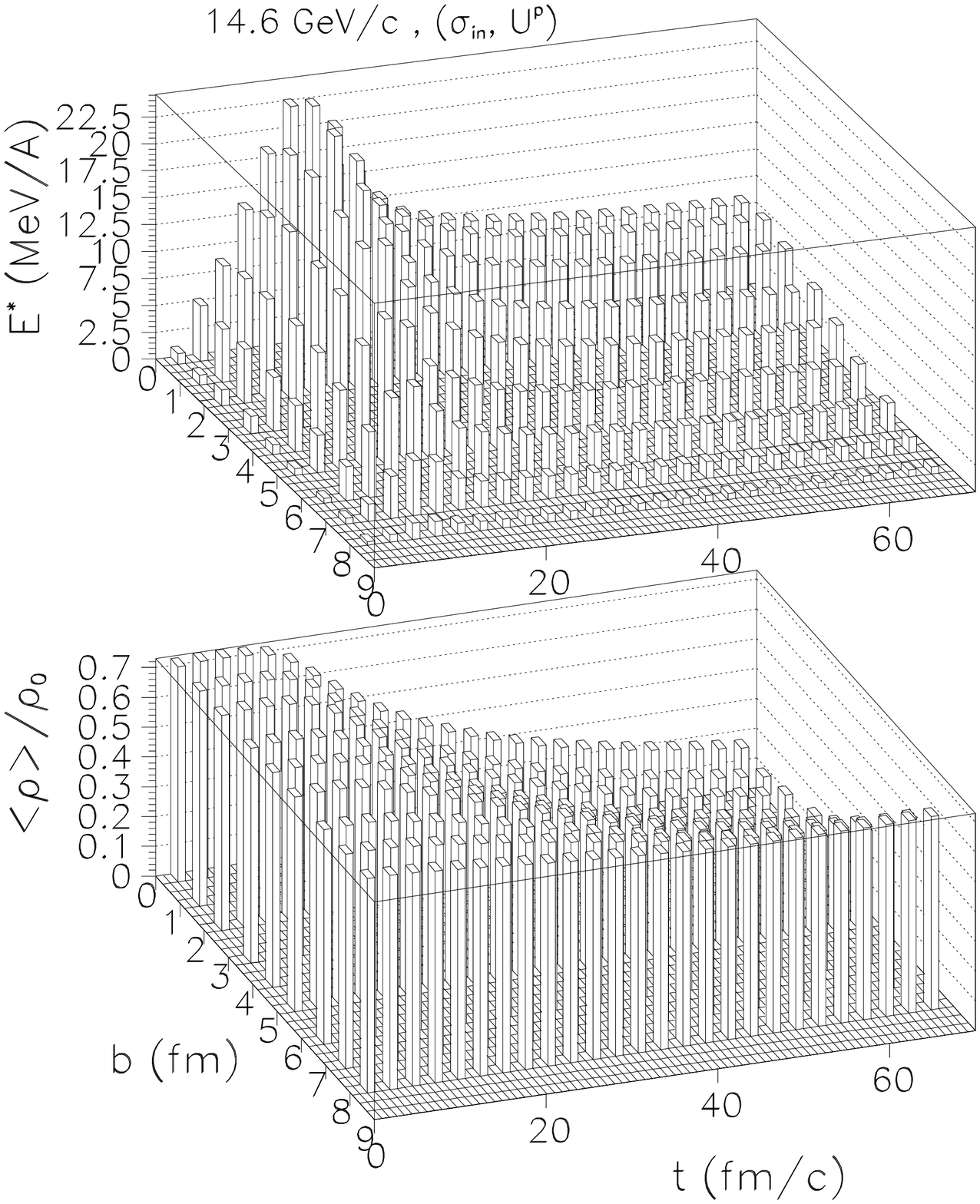,width=5.5in}}
\center{FIG. 10}
\end{figure}

\begin{figure}
\centerline{\psfig{figure=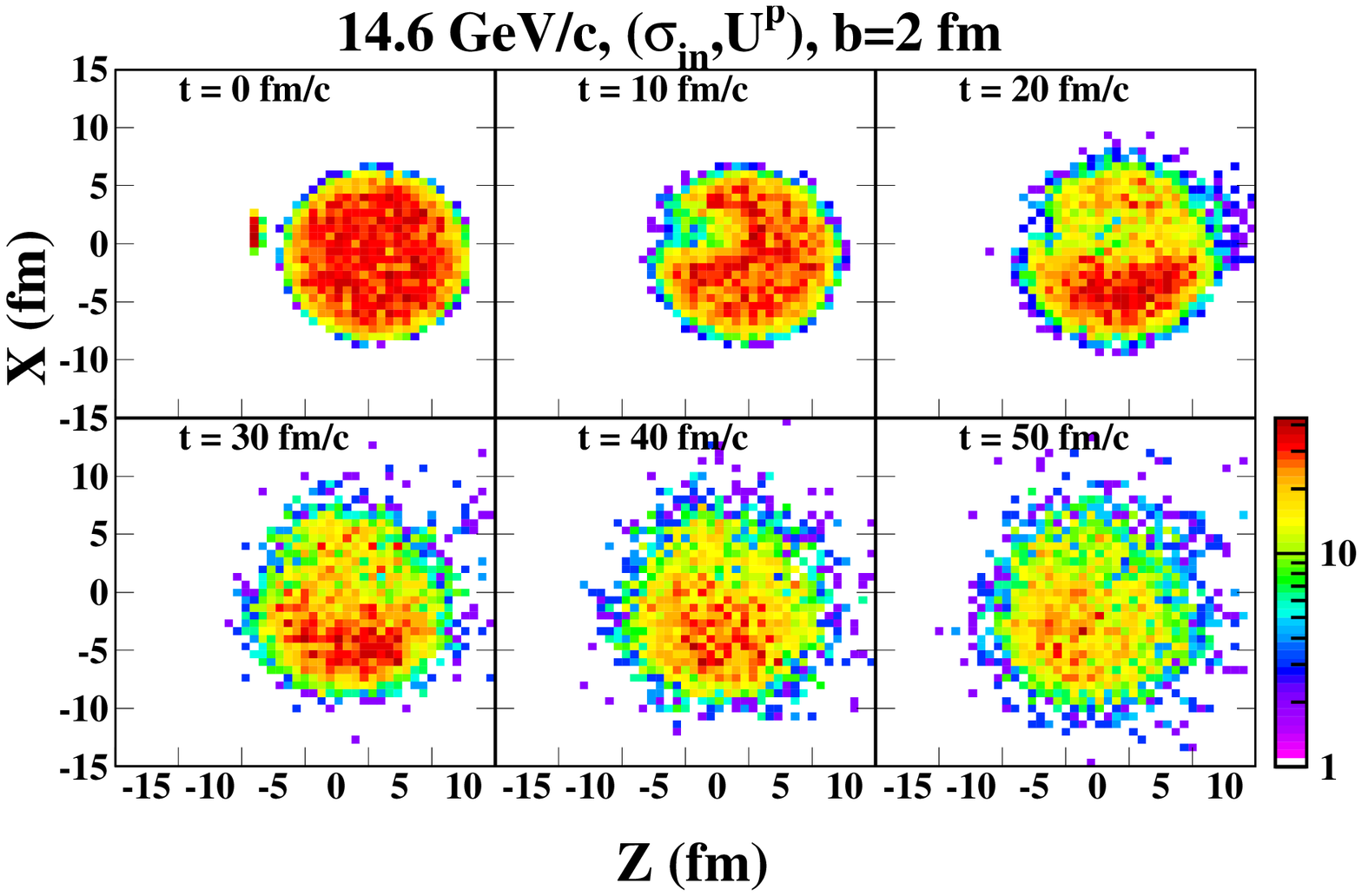,width=5.5in}}
\center{FIG. 11}
\end{figure}

\newpage
\vspace{2in}
\begin{figure}
\centerline{\psfig{figure=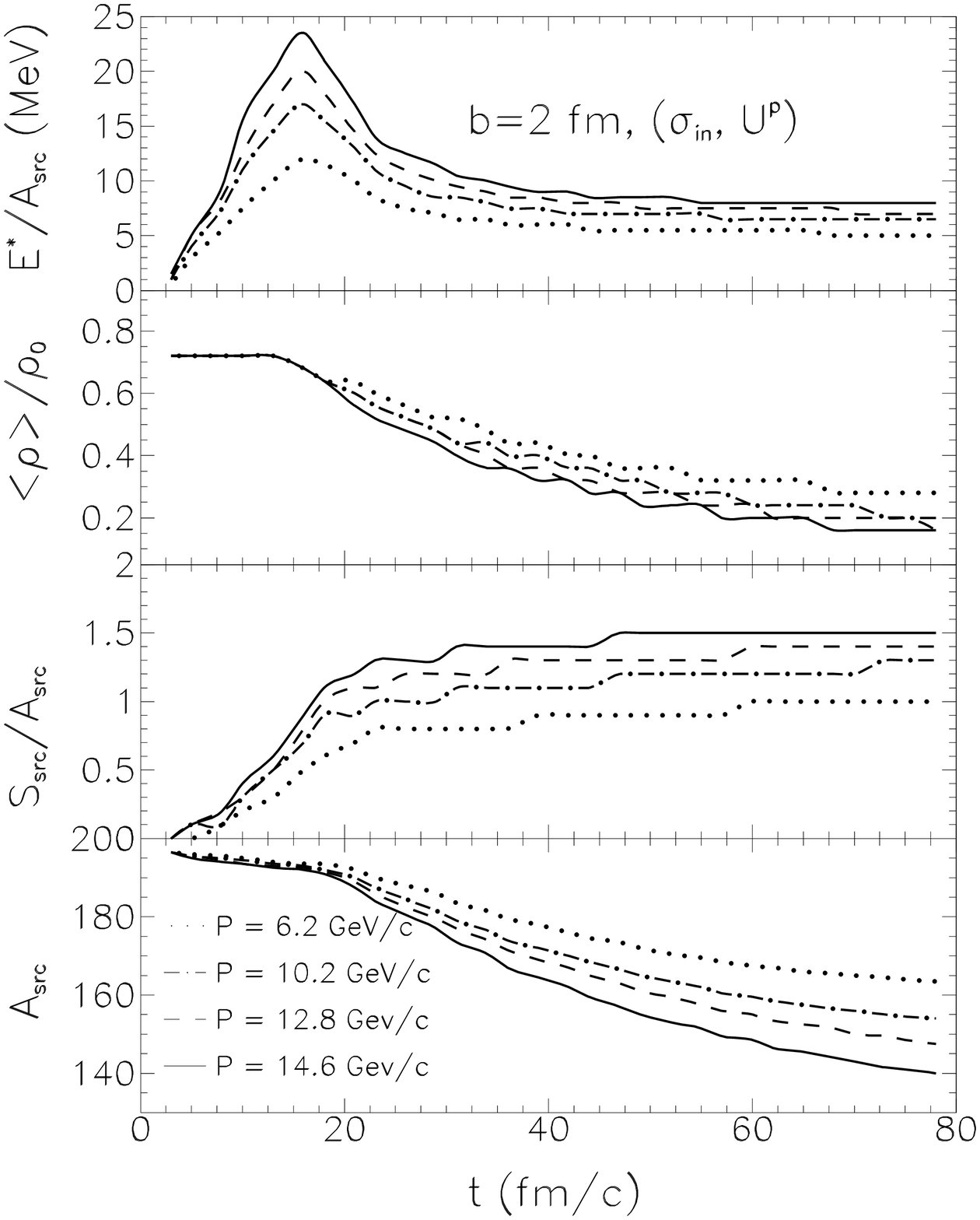,width=5.5in}}
\center{FIG. 12}
\end{figure}

\newpage
\begin{figure}
\centerline{\psfig{figure=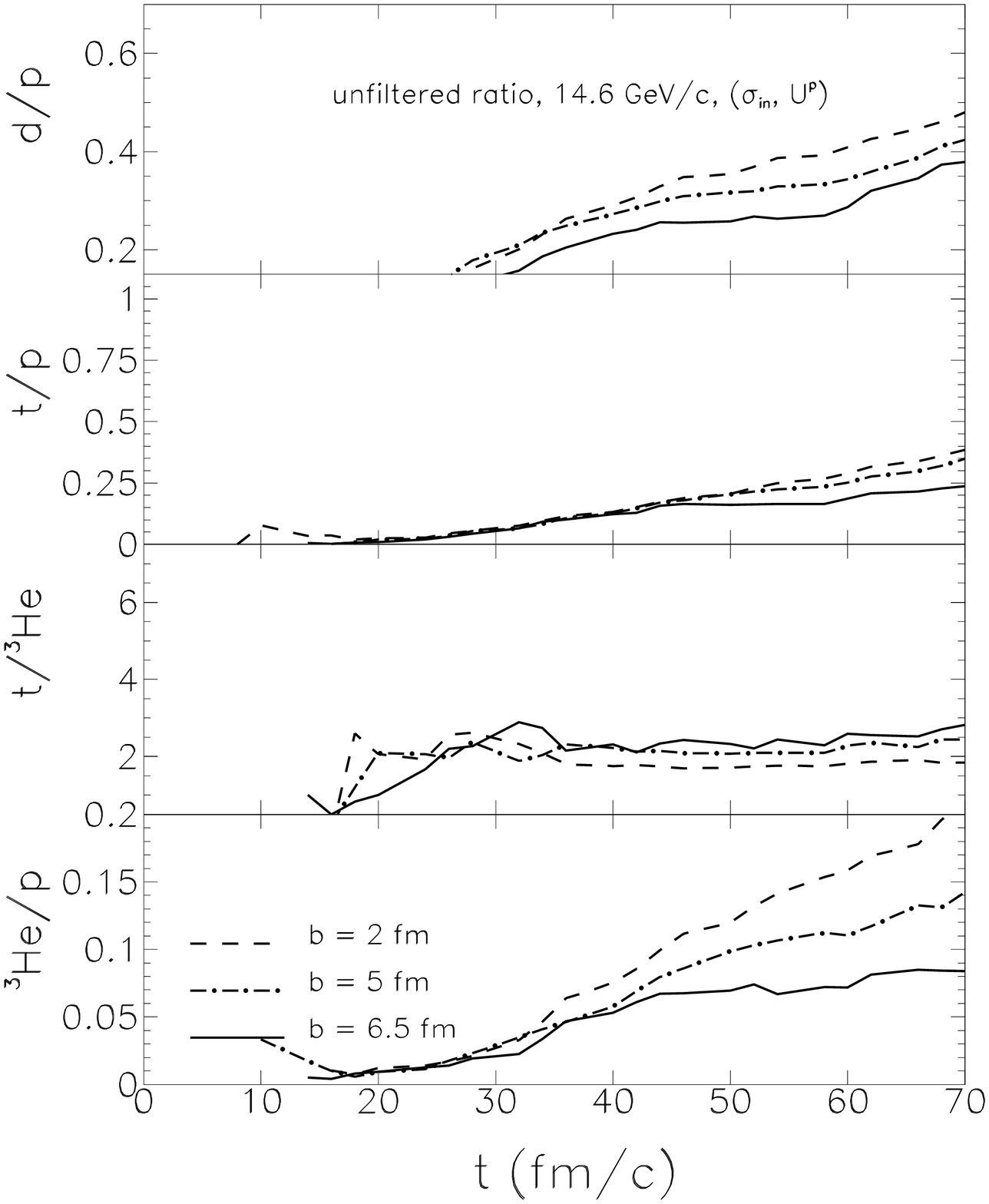,width=5.5in}}
\center{FIG. 13}
\end{figure}

\newpage
\begin{figure}
\centerline{\psfig{figure=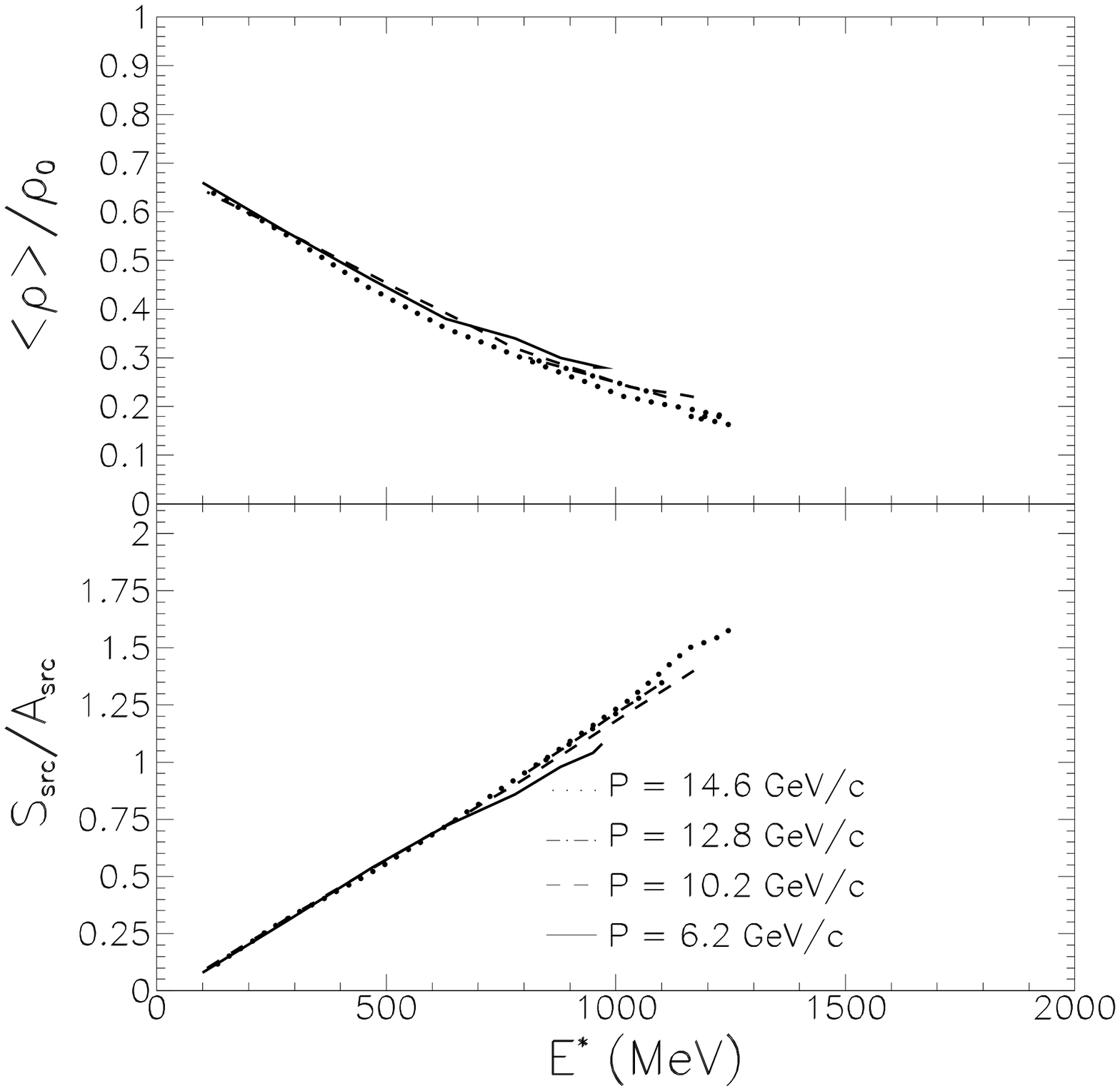,width=5.5in}}
\center{FIG. 14}
\end{figure}

\begin{thebibliography}{10}

\bibitem{beau2}
L. Beaulieu {\em et al.},
\newblock {}
\newblock {\rm Phys. Rev. C {\bf 64}, 064604 (2001).}

\bibitem{hsi}
W.-C. Hsi {\em et al.},
\newblock {}
\newblock {\rm Phys. Rev. Lett. {\bf 79}, 817 (1997).}

\bibitem{gold}
F. Goldenbaum {\em et al.},
\newblock {}
\newblock {\rm Phys. Rev. Lett. {\bf 77}, 1230 (1996).}

\bibitem{Avd1}
S.P. Avdeyev {\em et al.},
\newblock {}
\newblock {\rm Euro. Phys. J. {\bf A3}, 75 (1998).}

\bibitem{lef1}
T. Lefort {\em et al.},
\newblock {}
\newblock {\rm Phys. Rev. Lett. {\bf 83}, 4033 (1999).}

\bibitem{beau}
L. Beaulieu {\em et al.},
\newblock {}
\newblock {\rm Phys. Lett. {\bf B463}, 159 (1999).}

\bibitem{lef2}
T. Lefort {\em et al.},
\newblock {}
\newblock {\rm Phys. Rev. C {\bf 62}, 031604 (2000).}

\bibitem{wang}
G. Wang, K. Kwiatkowski and V.E. Viola,
\newblock {}
\newblock {\rm Phys. Rev. C {\bf 53}, 1811 (1996).}


\bibitem{Jeukenne}
J. P. Jeukenne, A. Lejeune and C. Mahaux,
\newblock {}
\newblock {\rm Phys. Rep. {\bf 25}, 83 (1976).}

\bibitem{Zhang}
J. Zhang, S. Das Gupta and C. Gale,
\newblock {}
\newblock {\rm Phys. Rev. C {\bf 50}, 1617 (1994).}

\bibitem{Pan}
Q. Pan and P. Danielewicz,
\newblock {}
\newblock {\rm Phys. Rev. Lett. {\bf 70}, 2062 (1993).}

\bibitem{dan00}
P.\ Danielewicz,
\newblock {}
\newblock{\rm Nucl.\ Phys.\ {\bf A673}, 375 (2000);
 Acta Phys. Polon. B {\bf 33}, 45 (2002).}


\bibitem{isab}
Y. Yariv and Z. Fraenkel,
\newblock {}
\newblock {\rm Phys. Rev. C {\bf 26}, 2138 (1982).}

\bibitem{cugn}
J. Cugnon, D. Kinet and J. Vandermeulen,
\newblock {}
\newblock {\rm Nucl. Phys. {\bf A379}, 553 (1982); J. Cugnon,
\it{ibid} 462, 751 (1987).}

\bibitem{qgsm}
V.D. Toneev, N.S. Amelin, K.K. Gudima and S. Yu Sivoklokov,
\newblock {}
\newblock {\rm Nucl. Phys. {\bf A519}, 463 (1990).}

\bibitem{ilji}
A.S. Iljinov, V.I. Nazaruk and S.E. Chigrinov,
\newblock {}
\newblock {\rm Nucl. Phys. {\bf A382}, 378 (1982).}

\bibitem{dan}
P. Danielewicz and G.F. Bertsch,
\newblock {}
\newblock {\rm Nucl. Phys. {\bf A533}, 712 (1991).}


\bibitem{bauer}
W. Bauer, G.F. Bertsch, W. Cassing and U. Mosel,
\newblock {}
\newblock {\rm Phys. Rev. C {\bf 34}, 2127 (1984); W. Bauer, Phys. Rev.
Lett. {\bf 61}, 2534 (1988).}

\bibitem{bus75}
W. Busza et al.,
\newblock {}
\newblock {\rm Phys. Rev. Lett. {\bf 34}, 836 (1975).}

\bibitem{che99}
I. Chemakin et al.,
\newblock {}
\newblock {\rm Phys. Rev. C {\bf 60}, 024902 (1999).}

\bibitem{kwia}
K. Kwiatkowski,
\newblock {}
\newblock {\rm Nucl. Instr. Meth. {\bf A360}, 571 (1995).}

\bibitem{lef3}
T. Lefort {\em et al.},
\newblock {}
\newblock {\rm Phys. Rev. C {\bf 64}, 064603 (2001).}

\bibitem{Berk}
M. Kleine Berkenbusch {\em et al.},
\newblock {}
\newblock {\rm Phys. Rev. Lett. {\bf 88}, 022701 (2002).}

\bibitem{Ell1}
J.B. Elliot {\em et al.},
\newblock {}
\newblock {\rm Phys. Rev. Lett. {\bf 88}, 042701 (2002).}

\bibitem{smm}
J.P. Bondorf, A.S. Botvina, A.S. Iljinov, I.N. Mishustin, K. Sneppen,
\newblock {}
\newblock {\rm Phys. Rep. 257 (1995) 133.}


\bibitem{morley}
K.B. Morley {\em et al.},
\newblock {}
\newblock {\rm Phys. Rev. C {\bf 54}, 737 (1996).}

\end{thebibliography}
\end{document}